# Force Analysis for Interactions beyond the Closest Neighbor in a Periodic Structure

**Farhad Farzbod**
Assistant Professor, University of Mississippi
farzbod@olemiss.edu

**Onome E. Scott-Emuakpor**
Senior Aerospace Engineer
Aerospace Systems Directorate (AFRL/RQTI)

ABSTRACT

Periodic structures are a type of metamaterials in which their physical properties not only depend on the unit cell materials but also the way unit cells are arranged and interact with each other. Periodic structure have Interesting wave propagation properties making them suitable materials for acoustic filters and wave beaming devices. Bloch analysis is the main tool to analyze wave propagation in these structures. In most if not all engineering structures, each unit cell is interacting with adjacent cells. As such methods developed for vibrational and wave propagation analysis of engineering periodic structures, address forces exerted by the closest neighbor only. Since metamaterials properties depend on the interactions of unit cell with neighboring cells, more interactions means more complex band structure. In this paper, we address force analysis when interactions are beyond the closest neighbor. This analysis, lays the foundation for vibrational analysis of structures in which interactions are not restricted to the closest neighbor.

1. INTRODUCTION

Periodic structures and metamaterials have the interesting property that their vibrational characteristics are not only a function of their unit cell's materials and properties but also how the unit cells are interacting with one another. In most of the vibrational analysis of these materials Bloch's theorem [1] is used. Based on the Bloch's theorem, Mead et al. [2, 3] developed a method to investigate harmonic wave propagation in periodic structures. Since then, there have been several other researches in a wide range of applications [4-27]. This includes wave guides [4] and wave beaming [27-29], filters [1-3], nonlinear materials [8-10, 30, 31] and materials with dissipation [32-38]. In these works, the interaction between unit cells are limited to the nearest neighbor; i.e., another unit cell which shares some boundary coordinates with the unit cell under consideration. This is due to the fact that in most engineering structures, the interactions are limited to elastic forces. This is not the case for electromagnetic forces such as the interatomic forces in a crystal where each atom is interacting with atoms beyond its closest neighbor. Previously, we have investigated [39] the maximum number of the wavevectors at each frequency when the interaction between unit cells are beyond the nearest neighbor. We have showed that this maximum number depends on the number of interactions or the topology of interactions. In other words, we have showed that in order to achieve specific properties in the dispersion curves – in that case the number of wave vectors – the topology of interactions has to be changed. Structures in which electromagnetic elements, such as magnets, are integrated in the design can provide these types of interactions. The first step in designing such materials, is to analyze forces on each unit cells. This will provide a foundation for analysis of such systems similar to the analysis developed before for the nearest neighbor interactions [40-42].

## 2. BLOCH ANALYSIS AND PROBLEMATIC FORCE TREATMENT OF DISTANT NEIGHBORS

In this section we start by using Bloch analysis to study a very simple mass-spring system in which interactions are limited to the closest neighbor. We then consider interactions beyond the closest neighbor using the same method, which is proved to be erroneous. We then present the modified model and the generalization of the method. For the sake of simplicity, we assume that there is no damping in all the systems under consideration. If there is energy dissipation in the structure, we can modify all the subsequent procedures to include damping by the same technique explained in [33]. Also we should mention that any unit cell of a periodic structure, after invoking finite element analysis, can be modelled by a discrete system of equations. As such our focus here and in all subsequent sections will be on discrete structures.

For the unit cell of the simple mass spring system of Fig.1 we have:

$$\left(-\omega^2 \begin{bmatrix} m & 0 \\ 0 & 0 \end{bmatrix} + \begin{bmatrix} K_1 & -K_1 \\ -K_1 & K_1 \end{bmatrix}\right) \begin{bmatrix} q_1 \\ q_2 \end{bmatrix} = \begin{bmatrix} F_1 \\ F_2 \end{bmatrix}. \tag{1}$$

Following our previous technique [40] for the minimum set of displacements we have

$$\mathbf{q} = \begin{bmatrix} \tilde{\mathbf{q}} \\ \tilde{\mathbf{q}}_x \end{bmatrix} = \begin{bmatrix} q_1 \\ q_2 \end{bmatrix} = \mathbf{T}[q_1] = \begin{bmatrix} \mathbf{I} \\ \mathbf{T}_x \end{bmatrix} [q_1] = \begin{bmatrix} 1 \\ e^{2\pi i k} \end{bmatrix} [q_1], \tag{2}$$

in which *k* is the wave vector. Replacing eq. (2) into eq. (1) and multiplying both sides by $\mathbf{T}^H$:

$$[1 \quad e^{-2\pi i k}] \left(-\omega^2 \begin{bmatrix} m & 0 \\ 0 & 0 \end{bmatrix} + \begin{bmatrix} K_1 & -K_1 \\ -K_1 & K_1 \end{bmatrix}\right) \begin{bmatrix} 1 \\ e^{2\pi i k} \end{bmatrix} [q_1] = [1 \quad e^{-2\pi i k}] \begin{bmatrix} F_1 \\ F_2 \end{bmatrix}. \tag{3}$$

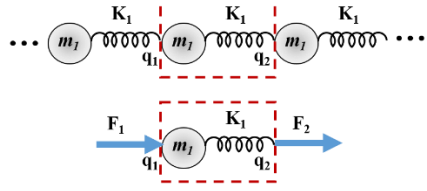

Fig. 1: A simple mass-spring chain. Forces on the boundary can be cancelled out by applying push-back operator.

As we showed previously, the right hand side is zero, so eq. (3) in this simple case becomes:

$$\left(-m\omega^2 + 2K_1 - K_1(e^{2\pi i k} + e^{-2\pi i k})\right) q_1 = 0. \tag{4}$$

Using Euler's formula and replacing $2\pi k$ by $\mu_x$ for the sake of simplicity, we get:

$$-m\omega^2 + 2K_1(1 - \cos(\mu_x)) = 0 \tag{5}$$

Now, we first try to use the same method for a simple system in which the interactions are beyond the closest neighbor. We start our investigation with the simplest case of single mass chain with interactions up to the second nearest neighbor, as the one depicted in Fig. 2. First we use simple calculation to find that the equation of motion for this mass-spring system should be:

$$\left(-2m\omega^2 + 4K_1 + 4K_2 - 2K_1(e^{i\mu_x} + e^{-i\mu_x}) - 2K_2(e^{i2\mu_x} + e^{-i2\mu_x})\right) q_1 = 0 \tag{6}$$

Next, we use the same method as before; writing the equation of motion for a unit cell and then applying push-forward and backward linear operator. The equation of motion for a unit cell of the system in Fig. 2 takes the form of:

$$\left(-\omega^2 \begin{bmatrix} m & 0 & 0 \\ 0 & m & 0 \\ 0 & 0 & 0 \end{bmatrix} + \begin{bmatrix} K_1 + K_2 & -K_1 & -K_2 \\ -K_1 & 2K_1 & -K_1 \\ -K_2 & -K_1 & K_1 + K_2 \end{bmatrix}\right) \begin{bmatrix} q_1 \\ q_2 \\ q_3 \end{bmatrix} = \begin{bmatrix} F_1 \\ F_2 \\ F_3 \end{bmatrix}. \tag{7}$$

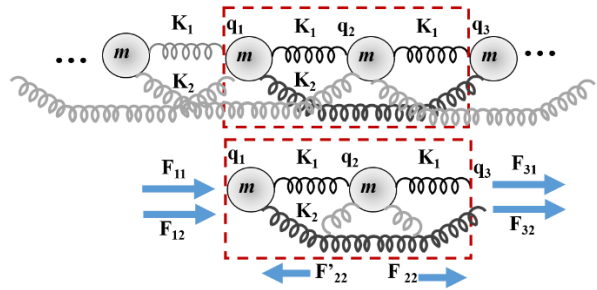

Fig. 2: (Top) A 1-D chain of masses in which masses interacts with the second closest neighbor as well as the closest neighbor. (Bottom) A cell which contains all the springs. The external forces would not cancel out by applying push-back operator in the same way they did for simple mass and spring chain.

Using push-forward and push-backward linear operators, the equation of motion for the unit cell becomes:

$$[1 \quad e^{-i\mu_x} \quad e^{-i2\mu_x}]\left(-\omega^2\begin{bmatrix}m & 0 & 0\\0 & m & 0\\0 & 0 & 0\end{bmatrix} + \begin{bmatrix}K_1+K_2 & -K_1 & -K_2\\-K_1 & 2K_1 & -K_1\\-K_2 & -K_1 & K_1+K_2\end{bmatrix}\right)\begin{bmatrix}1\\e^{i\mu_x}\\e^{i2\mu_x}\end{bmatrix}[q_1] =$$
$$[1 \quad e^{-i\mu_x} \quad e^{-i2\mu_x}]\begin{bmatrix}F_{11}+F_{12}\\F_{22}+F'_{22}\\F_{31}+F_{32}\end{bmatrix}. \tag{8}$$

It should be noted that $F_{12}$ is the external forces applied on the mass located at $q_1$ due to the spring with constant $K_2$. Left hand side of Eq. (8) can then be simplified as:

$$\left(-2m\omega^2 + 4K_1 + 2K_2 - 2K_1(e^{i\mu_x} + e^{-i\mu_x}) - K_2(e^{i2\mu_x} + e^{-i2\mu_x})\right) q_1, \tag{9}$$

which differs from the left hand side of eq. (6) and therefore does not equal to zero. This is due to the fact that unlike the other case with nearest neighbor interaction, the right hand side of eq. (8) would not vanish after applying the push-back operator, this is mainly due to the presence of external forces on the middle mass. We can get away from this problem of $F_{22}$ by including the other end of $K_2$ springs connected to the middle mass in the unit cell. However, this would cause other forces not to vanish, namely the ones that are now the middle ones with 'free' $K_2$ springs attached. It is therefore impossible to pick a "correct" unit cell in this case and apply the same technique as before. For this simple mass-spring system with no damping, it is straight forward to derive the equation of motion without having a framework or picking a unit cell. However, it is preferred to have a method with which we can generalize its applications to include all types of interactions, such as damping and interactions beyond 2nd or 3rd nearest neighbor. This is especially of interest since the main remarkable property of metametrials is that their properties do not solely depend on the unit cell properties, but rather depend on the way unit cells interact with each other. As a result, the more interactions we have, the more complex properties we can achieve.

The key to solving this problem is to treat forces of different layers of neighborhood – such as first, second or third nearest neighbors and so on – differently. Similar to the previous development, at the end, we are getting an eigenvalue equation in which the vector comprises of minimum number of displacements.

3. **TREATMENT OF FORCES FOR DISTANT NEIGHBORS**

Going back to the simplest case in which there are only closest neighbor interactions, we had [40]:

$$\bar{T}^T(-\omega^2 M + K)T\hat{q} = \bar{T}^T F, \tag{10}$$

in which the right hand side was shown to be zero [41]. For now, we keep the external force term in the right hand side intact and rewrite eq. (10) as:

$$\omega^2 \bar{\mathbf{T}}^T \mathbf{M} \mathbf{T} \hat{\mathbf{q}} = \bar{\mathbf{T}}^T \mathbf{K} \mathbf{T} \hat{\mathbf{q}} - \bar{\mathbf{T}}^T \mathbf{F}, \tag{11}$$

in which all the forces, including the spring forces, are on the right hand side. The left hand side which contains mass matrix, at the end of the matrix multiplication, would be the mass matrix corresponding to the minimum number of displacements. We rewrite the left hand side as $\omega^2 \hat{\mathbf{M}} \hat{\mathbf{q}}$.

Previously, in a 1-D chain structure we defined $\mathbf{T}$ and $\bar{\mathbf{T}}^T$ as linear push forward and backward operator for the closest neighbor. They contained the terms $e^{i\mu_x}$ and $e^{-i\mu_x}$. Similarly, we can define push forward and backward operator for the second nearest neighbor as $\mathbf{T}_2$ and $\bar{\mathbf{T}}_2^T$ in which $e^{i\mu_x}$ and $e^{-i\mu_x}$ are replaced by $e^{i2\mu_x}$ and $e^{-i2\mu_x}$. So the right hand side with the m$^{th}$ nearest neighbor would take the form of:

$$\bar{\mathbf{T}}_m^T \mathbf{K}_m \mathbf{T}_m \hat{\mathbf{q}} - \bar{\mathbf{T}}_m^T \mathbf{F} \tag{12}$$

in which $\mathbf{K}_m$ is the stiffness matrix comprising of stiffness elements which relates the unit cell to the m$^{th}$ nearest neighbor. Similarly, the second term in the eq. (12) corresponds to the boundary forces due to the interaction with the m$^{th}$ nearest neighbor. Following the same argument as the one in [41], the second term would be zero. Consequently the equation of motion in the case of interactions up to the n$^{th}$ nearest neighbor takes the form of:

$$\left(-\omega^2 \hat{\mathbf{M}} + \sum_{m=0,..n} \bar{\mathbf{T}}_m^T \mathbf{K}_m \mathbf{T}_m\right) \hat{\mathbf{q}} = 0 \tag{13}$$

Note that $K_1$ and $K_2$ are the stiffness matrices corresponding to the first and second closest neighbor. In order to facilitate spring forces inside the unit cell in a simplified formula, we denote this internal stiffness by matrix $K_0$ and we define $\mathbf{T}_0$ as the unity matrix $\mathbf{I}$. This will be further clarified through couple of examples. Revisiting the example structure of Fig. 2, forces can be separated as depicted in Fig. 3. For this structure, we can write:

$$\bar{\mathbf{T}}_1^T \mathbf{K}_1 \mathbf{T}_1 \hat{\mathbf{q}} - \bar{\mathbf{T}}_1^T \mathbf{F} = \begin{bmatrix} 1 & e^{-i\mu_x} \end{bmatrix} \left(\begin{bmatrix} K_1 & -K_1 \\ -K_1 & K_1 \end{bmatrix}\right) \begin{bmatrix} 1 \\ e^{i\mu_x} \end{bmatrix} [q_1] - \begin{bmatrix} 1 & e^{-i\mu_x} \end{bmatrix} \begin{bmatrix} F_{11} \\ F_{21} \end{bmatrix}, \tag{14}$$

$$\bar{\mathbf{T}}_2^T \mathbf{K}_2 \mathbf{T}_2 \hat{\mathbf{q}} - \bar{\mathbf{T}}_2^T \mathbf{F} = \begin{bmatrix} 1 & e^{-i2\mu_x} \end{bmatrix} \left(\begin{bmatrix} K_2 & -K_2 \\ -K_2 & K_2 \end{bmatrix}\right) \begin{bmatrix} 1 \\ e^{i2\mu_x} \end{bmatrix} [q_1] - \begin{bmatrix} 1 & e^{-i2\mu_x} \end{bmatrix} \begin{bmatrix} F_{12} \\ F_{32} \end{bmatrix}. \tag{15}$$

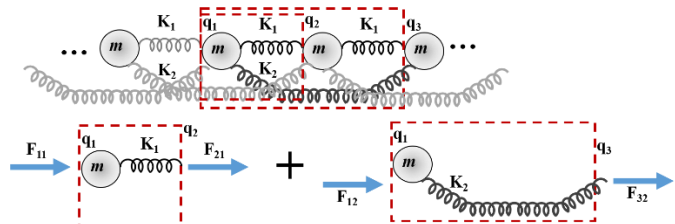

Fig. 3: (Top) A 1-D chain of masses. (Bottom) Interactions are categorized as 1$^{st}$ and 2$^{nd}$ closest neighbor interactions and they are dealt with differently using different push-back operators.

With similar procedure as the one in [41], we can show that the second terms on the right hand side of eq.'s (14) and (15) are net forces on a mass-less boundary and as such zero (forces are illustrated in Fig. 3) Therefore eq. (13) in this case can be written as:

$$\left(-\omega^2 m + \begin{bmatrix}1 & e^{-i\mu_x}\end{bmatrix}\begin{bmatrix}K_1 & -K_1 \\ -K_1 & K_1\end{bmatrix}\begin{bmatrix}1 \\ e^{i\mu_x}\end{bmatrix} + \begin{bmatrix}1 & e^{-i2\mu_x}\end{bmatrix}\begin{bmatrix}K_2 & -K_2 \\ -K_2 & K_2\end{bmatrix}\begin{bmatrix}1 \\ e^{i2\mu_x}\end{bmatrix}\right)[q_1] = 0, \quad (16)$$

which after multiplication is the same equation as eq. (6).

Next, we look at the system depicted in Fig. 4 with a more complex structure. First we consider internal spring forces in the unit cell. Displacements, push forward operator and the stiffness matrix can be written as:

$$[\tilde{q}] = T_0 \begin{bmatrix}q_1 \\ q_2 \\ q_3\end{bmatrix}, T_0 = I, K_0 = \begin{bmatrix}K_1 + K_4 & -K_1 & -K_4 \\ -K_1 & K_1 + K_2 & -K_2 \\ -K_4 & -K_2 & K_2 + K_4\end{bmatrix} \quad (17)$$

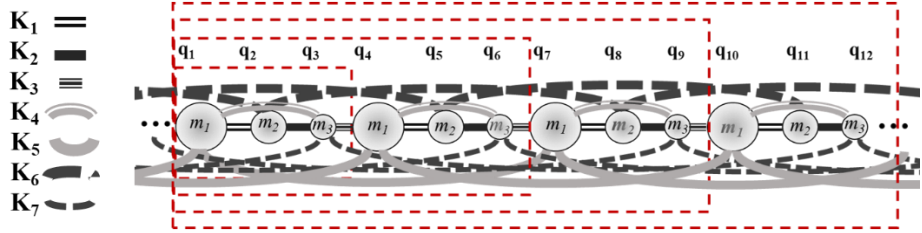

Fig. 4: A chain with three masses and interactions up to the 3$^{rd}$ closest neighbor.

Considering now the closest neighbor interactions, we first write displacement vector and the push forward operator as:

$$\begin{bmatrix}\tilde{q} \\ \tilde{q}_x\end{bmatrix} = \begin{bmatrix}q_1 \\ q_2 \\ q_3 \\ q_4 \\ q_5 \\ q_6\end{bmatrix} = T_1[\tilde{q}] = T_1 \begin{bmatrix}q_1 \\ q_2 \\ q_3\end{bmatrix}, T_1 = \begin{bmatrix}I \\ T_x\end{bmatrix} = \begin{bmatrix}1 & 0 & 0 \\ 0 & 1 & 0 \\ 0 & 0 & 1 \\ e^{i\mu_x} & 0 & 0 \\ 0 & e^{i\mu_x} & 0 \\ 0 & 0 & e^{i\mu_x}\end{bmatrix}. \quad (18)$$

For the stiffness matrix of the first nearest neighbor, we only consider springs that connects masses to the first nearest neighbor defined by displacement vector in eq. (18). In this example structure, only $m_3$ is connected to the $m_1$ of the nearest neighbor. As such the stiffness matrix takes the form of:

$$\mathbf{K_1} = \begin{bmatrix} 0 & 0 & 0 & 0 & 0 & 0 \\ 0 & 0 & 0 & 0 & 0 & 0 \\ 0 & 0 & K_3 & -K_3 & 0 & 0 \\ 0 & 0 & -K_3 & K_3 & 0 & 0 \\ 0 & 0 & 0 & 0 & 0 & 0 \\ 0 & 0 & 0 & 0 & 0 & 0 \end{bmatrix}. \quad (19)$$

Similarly for the second nearest neighbor we only consider springs that connects masses to the second nearest. In this case push forward operator, displacement vector and the corresponding stiffness matrix are defined as:

$$\begin{bmatrix} q_1 \\ q_2 \\ q_3 \\ q_7 \\ q_8 \\ q_9 \end{bmatrix} = \begin{bmatrix} 1 & 0 & 0 \\ 0 & 1 & 0 \\ 0 & 0 & 1 \\ e^{i2\mu_x} & 0 & 0 \\ 0 & e^{i2\mu_x} & 0 \\ 0 & 0 & e^{i2\mu_x} \end{bmatrix} \begin{bmatrix} q_1 \\ q_2 \\ q_3 \end{bmatrix}, \quad \mathbf{K_2} = \begin{bmatrix} K_5 & 0 & 0 & -K_5 & 0 & 0 \\ 0 & K_6 & 0 & 0 & -K_6 & 0 \\ 0 & 0 & 0 & 0 & 0 & 0 \\ -K_5 & 0 & 0 & K_5 & 0 & 0 \\ 0 & -K_6 & 0 & 0 & K_6 & 0 \\ 0 & 0 & 0 & 0 & 0 & 0 \end{bmatrix} \quad (20)$$

And for the third nearest neighbor:

$$\begin{bmatrix} q_1 \\ q_2 \\ q_3 \\ q_{10} \\ q_{11} \\ q_{12} \end{bmatrix} = \begin{bmatrix} 1 & 0 & 0 \\ 0 & 1 & 0 \\ 0 & 0 & 1 \\ e^{i3\mu_x} & 0 & 0 \\ 0 & e^{i3\mu_x} & 0 \\ 0 & 0 & e^{i3\mu_x} \end{bmatrix} \begin{bmatrix} q_1 \\ q_2 \\ q_3 \end{bmatrix}, \quad \mathbf{K_3} = \begin{bmatrix} 0 & 0 & 0 & 0 & 0 & 0 \\ 0 & 0 & 0 & 0 & 0 & 0 \\ 0 & 0 & K_7 & 0 & 0 & -K_7 \\ 0 & 0 & 0 & 0 & 0 & 0 \\ 0 & 0 & 0 & 0 & 0 & 0 \\ 0 & 0 & -K_7 & 0 & 0 & K_7 \end{bmatrix} \quad (21)$$

As such, for the example structure of Fig. 4, eq. (13) would be stated as:

$$\left( -\omega^2 \begin{bmatrix} m_1 & 0 & 0 \\ 0 & m_2 & 0 \\ 0 & 0 & m_3 \end{bmatrix} + \begin{bmatrix} K_1 + K_3 + K_4 + 2K_5 - K_5(e^{i2\mu_x} + e^{-i2\mu_x}) & -K_1 & -K_4 - K_3(e^{-i\mu_x}) \\ -K_1 & K_1 + K_2 + 2K_6 - K_6(e^{i2\mu_x} + e^{-i2\mu_x}) & -K_2 \\ -K_4 - K_3(e^{i\mu_x}) & -K_2 & K_2 + K_3 + K_4 + 2K_7 - K_7(e^{i3\mu_x} + e^{-i3\mu_x}) \end{bmatrix} \right) \begin{bmatrix} q_1 \\ q_2 \\ q_3 \end{bmatrix} \quad (22)$$

Dispersion curves for various values of spring constants are depicted in Fig. 5. As can be seen, interaction beyond the closest neighbor changes the dispersion curve in a way that is not possible to achieve with just closest neighbor interaction. Namely, the number of possible wave vectors at each frequency [39]. This change can be seen even at relatively small values of K for distant interactions. The change in band structure can possibly make the dispersion curve non monotonic or "flat" in a range and increase the chance of wider band gap.

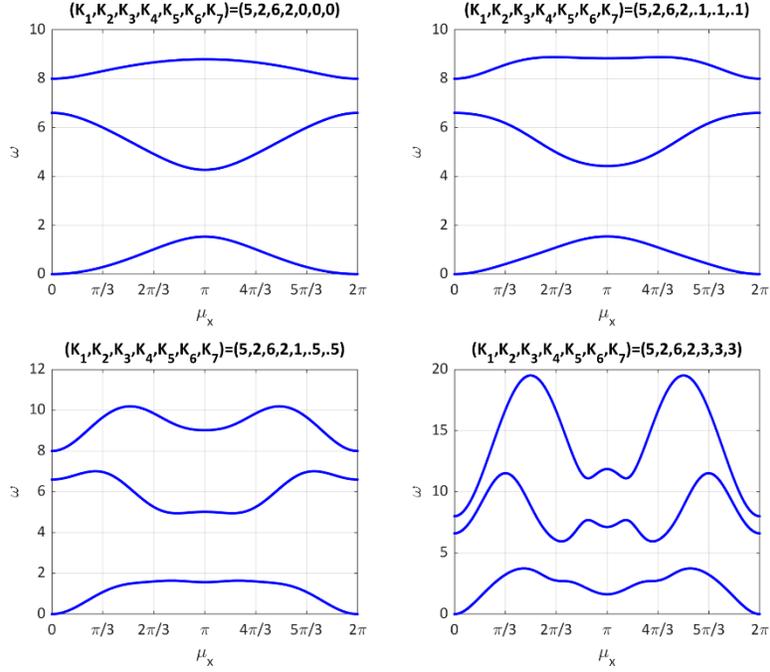

Fig. 5: Band structure of the three mass chain for various values of spring constants. As can be seen, dispersion curves are no longer monotonic in the interval of [0 π] when the interactions are beyond the closest neighbor.

To illustrate the application of this method, we next consider a 2D structure of Fig. 6. Since depicting interactions by springs beyond the closest neighbor makes the figure really busy, we tabulated the interactions in various tables. As can be seen in the Fig. 6, the cells are numbered in x and y direction. These numbers are reflected on the top left corner of the table to indicate interactions between the reference cell of (0,0) and the other cell. Masses on the first column and the first row represent the masses of the reference unit cell and the distanced unit cell, respectively. In this structure, we can write:

$$(-\omega^2 \hat{\mathbf{M}} + \sum_{q=0,1,2,\ r=0,1,2} \bar{\mathbf{T}}_{q,r}^T \mathbf{K}_{q,r} \mathbf{T}_{q,r}) \hat{\mathbf{q}} = \mathbf{0} \tag{23}$$

in which $\mathbf{T}_{q,r}$ is the push forward operator for the cell located at (q,r). For this structure, each unit cell has five masses. For the displacement vector and the internal springs, we have:

$$\hat{\mathbf{q}} = \begin{bmatrix} q_1 \\ q_2 \\ q_3 \\ q_4 \\ q_5 \end{bmatrix}, \mathbf{K}_{0,0} = \begin{bmatrix} K_1 + K_4 & -K_1 & 0 & -K_4 & 0 \\ -K_1 & K_1 + K_2 & -K_2 & 0 & 0 \\ 0 & -K_2 & K_2 + K_3 & -K_3 & 0 \\ -K_4 & 0 & -K_3 & K_3 + K_4 + K_8 & -K_8 \\ 0 & 0 & 0 & -K_8 & K_8 \end{bmatrix}. \tag{24}$$

As there are couple of interactions between the reference unit cell and the neighboring cells, here we only demonstrate formulas and mathematical relations for an example interactions. For this purpose we select the unit cell located at (2,1). Displacements of the reference unit cell and the cell located at (2,1) are stated as:

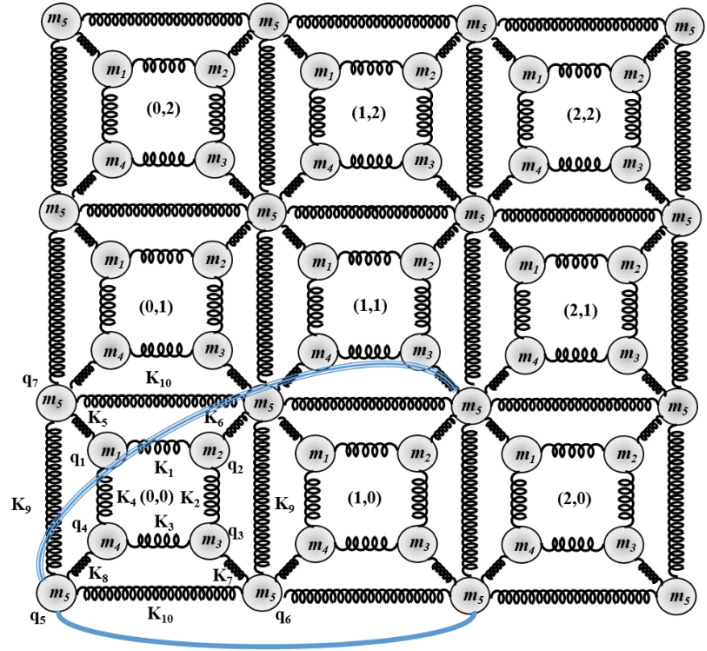

Fig. 6: A 2-D structure with five masses in a unit cell. The spring constants are tabulated in Table 1. Springs connecting distant masses are not depicted with the exception of $K_{15}$ (solid blue line) and $K_{19}$ (triple blue lines).

$$\begin{bmatrix} q_1 \\ q_2 \\ q_3 \\ q_4 \\ q_5 \\ q_{(2,1),1} \\ q_{(2,1),2} \\ q_{(2,1),3} \\ q_{(2,1),4} \\ q_{(2,1),5} \end{bmatrix} = \begin{bmatrix} 1 & 0 & 0 & 0 & 0 \\ 0 & 1 & 0 & 0 & 0 \\ 0 & 0 & 1 & 0 & 0 \\ 0 & 0 & 0 & 1 & 0 \\ 0 & 0 & 0 & 0 & 1 \\ e^{i(2\mu_x+\mu_y)} & 0 & 0 & 0 & 0 \\ 0 & e^{i(2\mu_x+\mu_y)} & 0 & 0 & 0 \\ 0 & 0 & e^{i(2\mu_x+\mu_y)} & 0 & 0 \\ 0 & 0 & 0 & e^{i(2\mu_x+\mu_y)} & 0 \\ 0 & 0 & 0 & 0 & e^{i(2\mu_x+\mu_y)} \end{bmatrix} \begin{bmatrix} q_1 \\ q_2 \\ q_3 \\ q_4 \\ q_5 \end{bmatrix}, \quad (25)$$

in which $q_{(2,1),1}$ is the displacement of $m_1$ of the unit cell located at (2,1). For the $K_{2,1}$ which represents the stiffness matrix due to the springs connecting the reference unit cell and the cell located at (2,1) we write:

$$\mathbf{K}_{2,1} = \begin{bmatrix} K_{16}+K_{17} & 0 & 0 & 0 & 0 & -K_{16} & 0 & 0 & -K_{17} & 0 \\ 0 & 0 & 0 & 0 & 0 & 0 & 0 & 0 & 0 & 0 \\ 0 & 0 & 0 & 0 & 0 & 0 & 0 & 0 & 0 & 0 \\ 0 & 0 & 0 & K_{18} & 0 & 0 & 0 & 0 & -K_{18} & 0 \\ 0 & 0 & 0 & 0 & K_{19} & 0 & 0 & 0 & 0 & -K_{19} \\ -K_{16} & 0 & 0 & 0 & 0 & K_{16} & 0 & 0 & 0 & 0 \\ 0 & 0 & 0 & 0 & 0 & 0 & 0 & 0 & 0 & 0 \\ 0 & 0 & 0 & 0 & 0 & 0 & 0 & 0 & 0 & 0 \\ -K_{17} & 0 & 0 & -K_{18} & 0 & 0 & 0 & 0 & K_{17}+K_{18} & 0 \\ 0 & 0 & 0 & 0 & -K_{19} & 0 & 0 & 0 & 0 & K_{19} \end{bmatrix}. \quad (26)$$

The remaining eight terms in eq. (23) can be found in the same manner. After substituting all the terms, dispersion surfaces are calculated. As explained in our previous paper [40] the boundaries of irreducible Brillouin zone does not necessarily include all the extremum of dispersion surfaces. This is more so in this highly asymmetrical structure. However, in order to show dispersion curves in a graph, we find ω's in eq. (23) for three lines of ($\mu_x$, $\mu_y$) = (0:π, 0), ($\mu_x$, $\mu_y$) = (π, 0:π) and $\mu_x$ = $\mu_y$ from π:0. These curves are depicted in Fig. 7. As can be observed, the curves show non monotonic behavior along each line. Dispersion surfaces for this structure are graphed in Fig. 8 and in Fig. 9 the surface for the third ω is shown. As evident in this figure, the maximum value happens in an off-diagonal position which is not on the boundary of the irreducible Brillouin zone.

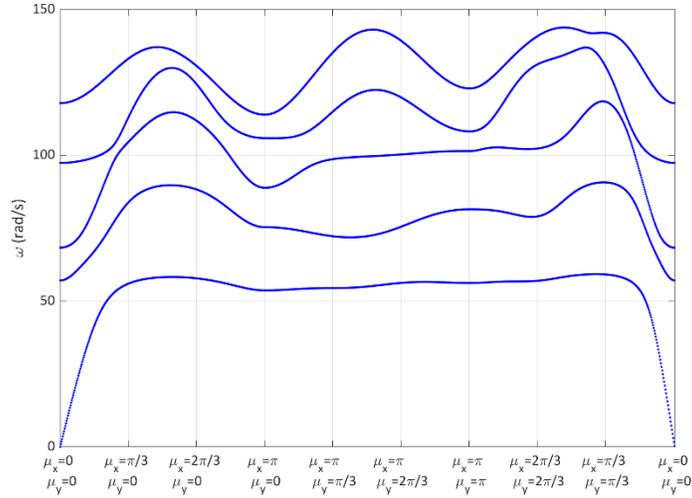

Fig. 7: Band structure of the five-mass 2-D structure when wave vectors are along three lines of ($\mu_x$, $\mu_y$) = (0:π, 0), ($\mu_x$, $\mu_y$) = (π, 0:π) and $\mu_x$ = $\mu_y$ from π:0.

## 4. CONCLUDING REMARKS

The complexity of metamaterials behavior, to a large extend, correlated to the complexity of the interactions between its unit cells. As such, interactions beyond the closest neighbors can lead to more complex and interesting metamaterials. In this paper, we provided a framework for analyzing forces when interactions are beyond the closest neighbor. This analysis paves the way for designing structures with more complex band structures.

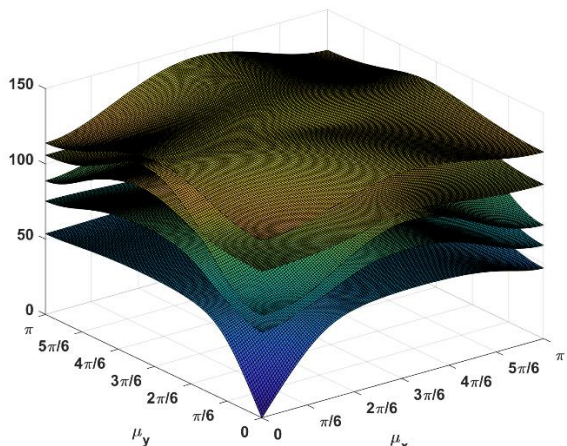

Fig. 8: Band structure of the of the five-mass 2-D structure.

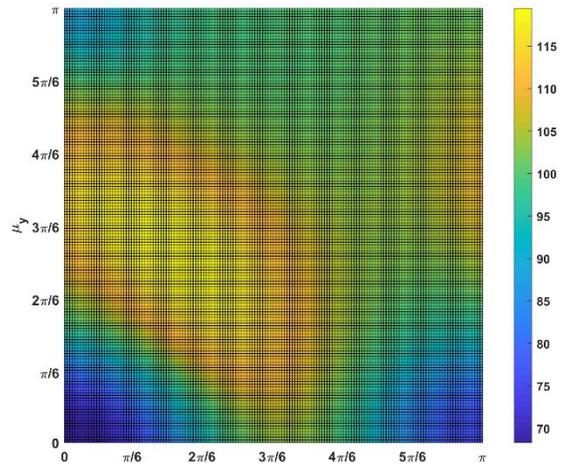

Fig. 9: Heat map of the third angular frequency of the five-mass 2-D structure.

Table 1: Spring constants

| (0,0) | $m_1$ | $m_2$ | $m_3$ | $m_4$ | $m_5$ |
|---|---|---|---|---|---|
| $m_1$ | 0 | $K_1$ | 0 | $K_4$ | 0 |
| $m_2$ |   | 0 | $K_2$ | 0 | 0 |
| $m_3$ |   |   | 0 | $K_3$ | 0 |
| $m_4$ |   |   |   | 0 | $K_8$ |
| $m_5$ |   |   |   |   | 0 |

| (1,0) | $m_1$ | $m_2$ | $m_3$ | $m_4$ | $m_5$ |
|---|---|---|---|---|---|
| $m_1$ | 0 | 0 | 0 | 0 | 0 |
| $m_2$ |   | 0 | 0 | 0 | 0 |
| $m_3$ |   |   | 0 | 0 | $K_7$ |
| $m_4$ |   |   |   | 0 | 0 |
| $m_5$ |   |   |   |   | $K_{10}$ |

| (0,1) | $m_1$ | $m_2$ | $m_3$ | $m_4$ | $m_5$ |
|---|---|---|---|---|---|
| $m_1$ | 0 | 0 | 0 | 0 | $K_5$ |
| $m_2$ |   | 0 | 0 | 0 | 0 |
| $m_3$ |   |   | 0 | 0 | 0 |
| $m_4$ |   |   |   | 0 | 0 |
| $m_5$ |   |   |   |   | $K_9$ |

| (1,1) | $m_1$ | $m_2$ | $m_3$ | $m_4$ | $m_5$ |
|---|---|---|---|---|---|
| $m_1$ | 0 | 0 | 0 | 0 | 0 |
| $m_2$ |   | 0 | 0 | 0 | $K_6$ |
| $m_3$ |   |   | 0 | 0 | 0 |
| $m_4$ |   |   |   | 0 | 0 |
| $m_5$ |   |   |   |   | 0 |

| (2,0) | $m_1$ | $m_2$ | $m_3$ | $m_4$ | $m_5$ |
|---|---|---|---|---|---|
| $m_1$ | $K_{11}$ | 0 | 0 | $K_{12}$ | 0 |
| $m_2$ |   | 0 | 0 | 0 | 0 |
| $m_3$ |   |   | 0 | 0 | 0 |
| $m_4$ |   |   |   | $K_{13}$ | $K_{14}$ |
| $m_5$ |   |   |   |   | $K_{15}$ |

| (2,1) | $m_1$ | $m_2$ | $m_3$ | $m_4$ | $m_5$ |
|---|---|---|---|---|---|
| $m_1$ | $K_{16}$ | 0 | 0 | $K_{17}$ | 0 |
| $m_2$ |   | 0 | 0 | 0 | 0 |
| $m_3$ |   |   | 0 | 0 | 0 |
| $m_4$ |   |   |   | $K_{18}$ | 0 |
| $m_5$ |   |   |   |   | $K_{19}$ |

| (0,2) | $m_1$ | $m_2$ | $m_3$ | $m_4$ | $m_5$ |
|---|---|---|---|---|---|
| $m_1$ | $K_{20}$ | 0 | 0 | 0 | 0 |
| $m_2$ |   | 0 | 0 | 0 | 0 |
| $m_3$ |   |   | 0 | 0 | 0 |
| $m_4$ |   |   |   | $K_{21}$ | 0 |
| $m_5$ |   |   |   |   | $K_{22}$ |

| (1,2) | $m_1$ | $m_2$ | $m_3$ | $m_4$ | $m_5$ |
|---|---|---|---|---|---|
| $m_1$ | $K_{23}$ | 0 | 0 | $K_{24}$ | 0 |
| $m_2$ |   | 0 | 0 | 0 | 0 |
| $m_3$ |   |   | 0 | 0 | 0 |
| $m_4$ |   |   |   | 0 | $K_{25}$ |
| $m_5$ |   |   |   |   | 0 |